\documentstyle[12pt]{article}
\begin{document}
\baselineskip=14pt
\pagestyle{empty}
\begin{center}

{\bf Theoretical study of the Josephson current through a junction 
including ferromagnetic insulator}
\end{center}

\begin{center}
{Masashi Yamashiro, Nobukatsu Yoshida, Yukio Tanaka and 
Satoshi Kashiwaya$^{\ast}$}
\end{center}

\begin{center}
Graduate School of Science and Technology,  Niigata University, Ikarashi, 
Niigata 950-21, Japan 
\end{center}

\begin{center}
$^{\ast}$Electrotechnical Laboratory, Tsukuba, Ibaraki 305, Japan
\end{center}
\vspace{12pt}

\noindent
{\bf ABSTRACT}

\vspace{12pt}

\noindent
The properties of Josephson junction including a ferromagnetic 
insulating layer are studied theoretically. 
The calculated current shows the anomalous dependence on the magnitude 
of the exchange interaction at the interface. 
The Josephson junction changes from 0-junction to $\pi$-junction 
with the increase of the magnitude of the exchange interaction. 

\vspace{12pt}


\vspace{12pt}

\noindent
{\bf INTRODUCTION}

\vspace{12pt}

\noindent
It is well known that the properties of the Josephson junction are 
strongly influenced  by the bound states 
formed around the insulator [1-5]. 
This bound states are current carrying states the  origins 
of which are multiple  Andreev reflection [6].
Previously, Furusaki and Tsukada derived a unified formula of the 
Josephson current, which is written by the coeficeints of the 
Andreev reflection [1,2]. 
This formula was extended to the Josephson junction including 
anisotropic superconductors [3-5]. 
On the other hand, 
Tanaka and Kashiwaya developed a theory of 
superconductor / 
ferromagnetic-insulator / superconductor (S/Fi/S) junction [7] 
based on the Furusaki's formula. 
They showed that the Josephson junction changes from 0-junction to 
$\pi$-junction with the increase of the 
magnitude of the exchange interaction strength. 
In this paper, we will 
calculate the Josephson current in S/Fi/S junction 
taking into account the exchange potential as well as ordinary 
Hartree potential 
at the interface which was ignored in the previous paper [7]. 
As in the previous results, 
the Josephson current changes sign 
with the increase of exchange interaction strength for 
fixed phases difference between two superconductors. 
Our results inclueds several existing 
 results as limiting cases [8-10].

\vspace{12pt}

\noindent
{\bf FORMULA AND RESULTS}

\vspace{12pt}

\noindent
We assume the system in the clean limit with perfectly flat interface 
which is perpendicular to the $x$-axis and is located at $x$=0. 
The Fermi wave number $k_{F}$ and the effective mass $m$ are assumed to be 
equal both on the left (negative $x$ side) and the right (positive $x$ side) 
superconductors. 
All dynamical spin processes are neglected and the magnetic influence 
on the superconductivity is limited to that through the exchange field 
in the Fi region. 
Since the momentum component parallel to the interface is conserved, 
the Bogoliubov-de Gennes equation can be expressed as 
\begin{equation}
\left( \begin{array}{cc}
h_{1,\sigma}(\theta) & \Delta(x) \\[12pt]
\Delta^{\ast}(x) & -h_{2,\sigma}(\theta)
\end{array} \right)
\Psi_{\sigma}(x,\theta)=E\Psi_{\sigma}(x,\theta),\hspace{18pt}
\sigma=\uparrow,\hspace{8pt} \downarrow,
\label{eq1}
\end{equation}

\begin{equation}
h_{1(2),\sigma}(\theta)=
\displaystyle
-\frac{\hbar^{2}}{2m}
\displaystyle
\frac{d^{2}}{dx^{2}}+(H_{i}-(+)\rho_{\sigma}H_{m})\delta(x)-
\displaystyle
\frac{\hbar^{2}k^{2}_{F}\cos^{2}\theta}{2m},\hspace{16pt}
\rho_{\uparrow}=1,\hspace{10pt}\rho_{\downarrow}=-1.
\label{eq2}
\end{equation}
In the above, the wave function $\Psi_{\sigma}(x,\theta)$ expresses 
the process where an electron-like quasiparticle is injected from 
the left superconductor. 
$H_{i}$ and $H_{m}$ represent the ordinary barrier potential and 
exchange interaction, respectively.
The quantity $\theta$ denotes the injection angle of the quasiparticle 
which is measured from the normal to the interface. 
The energy of quasiparticles $E$ is measured from the Fermi energy $E_{F}$.
We assume that the pair potential is expressed as 
$\Delta(x)$=$\Delta(T)e^{i\phi_{L}}$ for $x<0$, and 
$\Delta(x)$=$\Delta(T)e^{i\phi_{R}}$ for $x>0$.
The Andreev reflection coefficient $a_{\sigma}(\phi,\theta)$ is obtained 
as follows.
\begin{equation}
a_{\sigma}(\phi,\theta)=
\displaystyle
\frac{\Delta(T)\left[-E\sin^{2}(\phi/2)+i\Omega\left\{\sin(\phi/2)\cos(\phi/2)
+\rho_{\sigma}z_{m\theta}\right\}\right]}
{E^{2}(1+Z_{\theta}^{2})-
\Delta^{2}(T)\left\{\cos^{2}(\phi/2)+Z^{2}_{\theta}\right\}-
2iE \Omega \rho_{\sigma}z_{m\theta}}.
\label{eq3}
\end{equation}
In the Eq.(\ref{eq3}), $\Omega$=$\sqrt{E^{2}-\Delta^{2}(T)}$. 
The phase difference between the 
left and right superconductor is denoted as 
$\phi$=$\phi_{L}-\phi_{R}$. 
Barrier parameters at the interface are given by
\begin{equation}
\begin{array}{l}
z_{m\theta}=
\displaystyle
\frac{mH_{m}}{\hbar^{2}k_{F}\cos\theta},\hspace{20pt}
z_{i\theta}=
\displaystyle
\frac{mH_{i}}{\hbar^{2}k_{F}\cos\theta} \\[24pt]
Z_{\theta}^{2}=z_{i\theta}^{2}-z_{m\theta}^{2}.
\end{array}
\label{eq4}
\end{equation}
Following the formulation in Ref.[1,2], the Josephson current is 
obtained using the generalized coefficient of the Andreev reflection 
$\bar{a}_{\sigma}(\phi,\theta)$ which is obtained  by analytic continuation 
of $E$ to $i\omega_{n}$ in $a_{\sigma}(\phi,\theta)$. 
The obtained result for a fixed injection angle $I(\phi,\theta)$ 
is expressed as 
\begin{equation}
I(\phi,\theta)=
\displaystyle
\frac{e\Delta(T)k_{B}T}{\hbar}
\displaystyle
\sum_{\omega_{n}}
\displaystyle
\frac{\bar{a}_{\uparrow}(\phi,\theta)+\bar{a}_{\downarrow}(\phi,\theta)-
\bar{a}_{\uparrow}(-\phi,\theta)-\bar{a}_{\downarrow}(-\phi,\theta)}
{2\Omega_{n}},
\label{eq5}
\end{equation}
with $\omega_{n}$=$2\pi k_{B}T(n+1/2)$ and 
$\Omega_{n}$=$\sqrt{\omega_{n}^{2}+\Delta^{2}(T)}$.
Performing the summation of the Matsubara frequency, 
the whole Josephson current $J(\phi)$ is obtained as 
\begin{equation}
\begin{array}{l}
\displaystyle
J(\phi)=\frac{k^{2}_{F}S}{\pi}\int_{0}^{\pi/2}
I(\phi,\theta)\cos\theta\sin\theta d\theta, \\
\noalign{\vskip 2ex}
I(\phi,\theta)=
\displaystyle
\frac{e\Delta^{2}(T)\sin\phi}
{2\hbar\left\{(1+Z_{\theta}^{2})^{2}+4z_{m\theta}^{2}\right\}}
\left\{
\displaystyle
\frac{\alpha(\phi,\theta)}{\varepsilon_{A}(\phi,\theta)}
\mbox{tanh} \displaystyle \frac{\varepsilon_{A}(\phi,\theta)}{2k_{B}T}+
\frac{\beta(\phi,\theta)}{\varepsilon_{B}(\phi,\theta)}
\mbox{tanh} \displaystyle \frac{\varepsilon_{B}(\phi,\theta)}{2k_{B}T}
\right \}
\end{array}
\label{eq6}
\end{equation}
with following quantities
\begin{equation}
\begin{array}{l}
\varepsilon_{A}^{2}(\phi,\theta)=\Delta^{2}(T) \\
\displaystyle
\times \frac{(1+Z_{\theta}^{2})(\cos^{2}(\phi/2)+Z_{\theta}^{2})+
2z_{m\theta}\left\{z_{m\theta}+\sqrt{z_{m\theta}^{2}+
(\cos^{2}(\phi/2)+Z_{\theta}^{2})\sin^{2}(\phi/2)}\right\}}
{(1+Z_{\theta}^{2})^{2}+4z_{m\theta}^{2}} 
\end{array}
\end{equation}
\begin{equation}
\begin{array}{l}
\varepsilon_{B}^{2}(\phi,\theta)=\Delta^{2}(T) \\
\displaystyle
\times \frac{(1+Z_{\theta}^{2})(\cos^{2}(\phi/2)+Z_{\theta}^{2})+
2z_{m\theta}\left\{z_{m\theta}-\sqrt{z_{m\theta}^{2}+
(\cos^{2}(\phi/2)+Z_{\theta}^{2})\sin^{2}(\phi/2)}\right\}}
{(1+Z_{\theta}^{2})^{2}+4z_{m\theta}^{2}} 
\end{array}
\end{equation}

\begin{equation}
\begin{array}{l}
\alpha(\phi,\theta)=\displaystyle \frac12
\left\{ (1+Z_{\theta}^{2})-
\displaystyle
\frac{z_{m\theta}(\cos\phi+Z_{\theta}^{2})}
{\sqrt{z_{m\theta}^{2}+(\cos^{2}(\phi/2)+Z_{\theta}^{2})\sin^{2}(\phi/2)}}
\right\} 
\end{array}
\end{equation}

\begin{equation}
\begin{array}{l}
\beta(\phi,\theta)=\displaystyle \frac12
\left\{ (1+Z_{\theta}^{2})+
\displaystyle
\frac{z_{m\theta}(\cos\phi+Z_{\theta}^{2})}
{\sqrt{z_{m\theta}^{2}+(\cos^{2}(\phi/2)+Z_{\theta}^{2})\sin^{2}(\phi/2)}}
\right\}.
\end{array}
\end{equation}
The quantity $S$ expresses the magnitude of the section of the junction. 
Figs.(\ref{fig1}) and 
(\ref{fig2}) show the calculated Josephson current $J(\phi)$ 
and maximum Josephson current $J_{C}$ for fixed $z_{i}$=1 and 
several values of $z_{m}$ ($z_{i(m)}$=$\cos\theta z_{i(m)\theta}$), 
which are normalized by the resistance in the normal state 
\begin{equation}
R_{N}^{-1}=
\displaystyle
\frac{e^{2}k_{F}^{2}S}{\pi^{2}\hbar}\int^{\pi/2}_{0}
\displaystyle
\frac{\cos\theta\sin\theta}{1+(z_{i\theta}-\rho_{\sigma}z_{m\theta})^{2}}
d\theta
\label{eq8}
\end{equation}
On the other hand, Figs.(\ref{fig3}) and (\ref{fig4}) are
corresponding figures of Figs.(\ref{fig1}) and (\ref{fig2}) 
for fixed $z_{m}$=1. 
As shown in Fig.(\ref{fig1}), 
when the magnitude of $z_{m}$ becomes larger than that of 
$z_{i}$, 
the Josephson junction changes from 0-junction to $\pi$-junction. 
For large $z_{m}$ or $z_{i}$, 
the temperature dependence of the 
maximum Josephson current $J_{C}$ [Figs.(\ref{fig2}),(\ref{fig4})] 
shows similar behavior to that found in the tunneling limit of 
the non-magnetic barrier by Ambegaokar and Baratoff [10].
For fixed $z_{m}$, with the increase of $z_{i}$, 
the Josephson junction changes from $\pi$-junction to 0-junction
[Fig.(\ref{fig3})]. 

\vspace{12pt}

\noindent
In this paper, we have developed a general formula of the d.c. 
Josephson current between the superconductor / 
ferromagnetic-insulator / superconductor junction 
taking into account 
both ordinary barrier potential as well as exchange interaction. 
The essence of the physics in our previous paper [7] 
does not change due to the existence of the ordinary  potential barrier. 
We hope the drastical change from 0 junction to $\pi$ junction will be 
observed experimentally near future. 

\vspace{12pt}

\noindent
{\bf REFERENCES}

\vspace{12pt}

\noindent
1. Furusaki A and Tsukada M  (1990) Physica {\bf C} 165\&166: 967-968

\noindent
2. Furusaki A and Tsukada M  (1991) Solid State Commun. 78: 299-302

\noindent
3. Tanaka Y and Kashiwaya S  (1996) Phys. Rev. {\bf B} 53: R11957-R11960

\noindent
4. Tanaka Y and Kashiwaya S  (1996) Phys. Rev. {\bf B} 53: 9371-9381

\noindent
5. Tanaka Y and Kashiwaya S  (1997) Phys. Rev. {\bf B} 56: 892-912

\noindent
6. Andreev AF  (1964) Sov. Phys. JETP 19: 1228-1231

\noindent
7. Tanaka Y and Kashiwaya S  (1997) Physica {\bf C} 274: 357-363

\noindent
8. Shiba H and Soda T (1969) Prog. Theor. Phys. 41: 25-44

\noindent
9. Kulik IO and Omel'yanchuk (1978) Sov. J. Low Temp. Phys. 4: 142-

\noindent
10. Ambegaokar V and Baratoff A (1963) Phys. Rev. Lett. 10: 486-489

\newpage
\noindent
\begin{figure}
{\bf FIGURE CAPTIONS}
\caption{Normalized $J(\phi)$ is plotted as a function of $\phi/\pi$,
with a:$z_{m}$=$0.5$, b:$z_{m}$=$1.5$, c:$z_{m}$=$2$ and d:$z_{m}$=$5$.
$z_{i(m)}$=$\cos\theta z_{i(m)\theta}$}
\label{fig1}

\caption{Normalized $J_{C}$ is plotted as a function of $T/T_{C}$,
with a:$z_{m}$=$0.5$, b:$z_{m}$=$1$, c:$z_{m}$=$2$ and d:$z_{m}$=$5$.
$z_{i(m)}$=$\cos\theta z_{i(m)\theta}$}
\label{fig2}

\caption{Normalized $J(\phi)$ is plotted as a function of $\phi/\pi$,
with a:$z_{i}$=$0.5$, b:$z_{i}$=$1.5$, c:$z_{i}$=$2$ and d:$z_{i}$=$5$.
$z_{i(m)}$=$\cos\theta z_{i(m)\theta}$}
\label{fig3}

\caption{Normalized $J_{C}$ is plotted as a function of $T/T_{C}$,
with a:$z_{i}$=$0.5$, b:$z_{i}$=$1$, c:$z_{i}$=$2$ and d:$z_{i}$=$5$.
$z_{i(m)}$=$\cos\theta z_{i(m)\theta}$}
\label{fig4}
\end{figure}

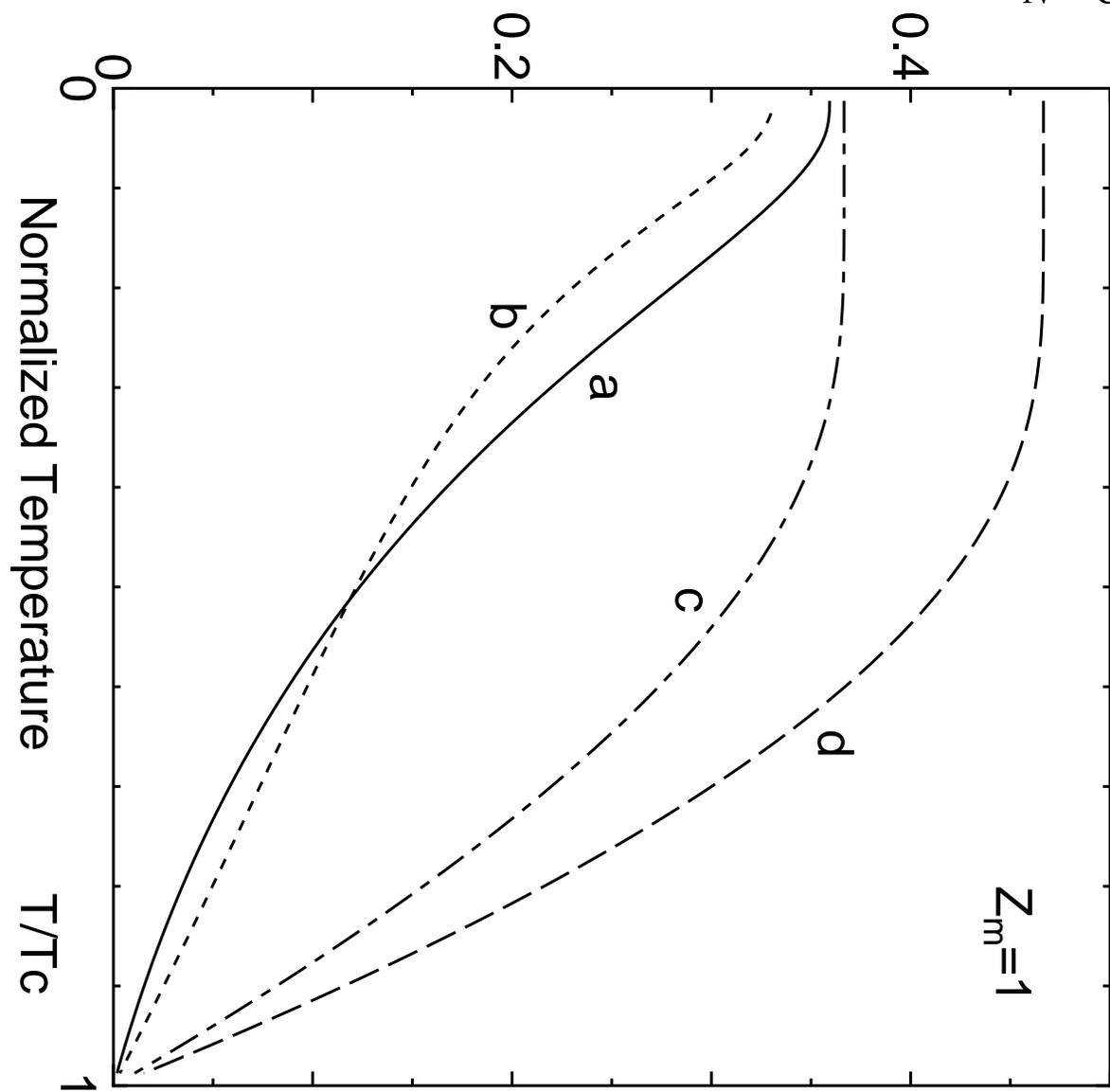
\end{document}